\documentclass[12pt,reqno]{amsproc}  

\usepackage{a4}
\usepackage[german,english]{babel}
\usepackage{amsmath}
\usepackage{amssymb}
\usepackage{graphicx}               

\theoremstyle{plain}            
\newtheorem{theorem}{Theorem}[section]
\newtheorem*{theorem*}{Theorem}
\newtheorem{proposition}[theorem]{Proposition}

\theoremstyle{definition}       

\theoremstyle{remark}

\numberwithin{equation}{section}

\DeclareMathOperator{\dom}    {dom}
\DeclareMathOperator{\spec}   {spec}

\newlength{\maxbreite}%
\newlength{\maxhoehe}%
\newlength{\maxtiefe}%

\newcommand{\stelldrueber}[3][0pt]{
  \settowidth{\maxbreite}{#3}%
  \settoheight{\maxhoehe}{#3}%
  \settodepth{\maxtiefe}{#2}%
  \addtolength{\maxhoehe}{\maxtiefe}%
  {\makebox[\maxbreite]{\raisebox{\maxhoehe}{\hspace{#1}#2}}%
  \makebox[0pt][r]{#3}}%
}

\newcommand{\overcirc}[1]       
{\stelldrueber[.45ex]{$\scriptscriptstyle\circ$}{${#1}$}}

\newcommand{\R}{\mathbb{R}} 
\newcommand{\C}{\mathbb{C}} 
\newcommand{\N}{\mathbb{N}} 
\newcommand{\Z}{\mathbb{Z}} 
\newcommand{\Sphere}{\mathbb{S}} 
\newcommand{\Torus}{\mathbb{T}} 
\newcommand{\eps}{\varepsilon} 
\renewcommand{\phi}{\varphi}   
\newcommand{\e}{\mathrm e}  
\newcommand{\im}{\mathrm i} 

\newcommand{\HS}{\mathcal H}                

\newcommand{\Ci} [1]{C^\infty ({#1})}       
\newcommand{\CiZ}[1]{C^\infty_\mrt eq. ({#1})}  
\newcommand{\Cci}[1]{C_{\mathrm c}^\infty ({#1})} 
\newcommand{\Lsqr}[1]{L_2({#1})}            
\newcommand{\lsqr}[1]{\ell_2({#1})}         
\newcommand{\Sob}[2][1]{\HS^{#1}({#2})}     
\newcommand{\SobZ}[2][1]{\HS^{#1}_\mrt eq.({#2})} 
\newcommand{\Sobn}[2][1]{{\overcirc {\mathcal H}{}^{#1}({#2})}}
\newcommand{\norm}[2][{}]{\|{#2}\|_{#1}}    
\newcommand{\normsqr}[2][{}]{\|{#2}\|^2_{#1}} 
\newcommand{\iprod}[3][{}]{\langle{#2},{#3}\rangle_{#1}}  
\newcommand{\bd}  {\partial}                

\newcommand{\restr}[1]{{\restriction}_{#1}} 
\newcommand{\map}[3]{{#1}\colon{#2}\longrightarrow{#3}} 
\newcommand{\set}[2]{\{ \, #1 \, | \, #2 \, \} } 

\usepackage{bbm}         
\newcommand{\1}{\mathbbm 1}                    

\newcommand{\id}[1][{}]{\mathrm{id_{#1}}}        

\reversemarginpar     

\newcommand{\Neu}{{-}}              
\newcommand{\Dir}{{+}}              
\newcommand{\DirNeu}{{\pm}}              

\newcommand{\laplacian}[1]{\Delta_{{#1}}}   
\newcommand{\laplacianD}[1]{\Delta^\Dir_{{#1}}}
\newcommand{\laplacianDN}[1]{\Delta^{\DirNeu} _{{#1}}}
\newcommand{\laplacianN}[1]{\Delta^\Neu_{{#1}}}
\newcommand{\laplacianZ}[1]{\Delta_{{#1}}(z)}

\newcommand{\EWD}[1]{\lambda^\Dir_{#1}}
\newcommand{\EWN}[1]{\lambda^\Neu_{#1}}
\newcommand{\EWZ}[1]{\lambda_{#1}(z)}  
\newcommand{\EWDN}[1]{\lambda^{\DirNeu}_{#1}}

\newcommand{\G}{\Gamma}               
\newcommand{\hG}{\widehat \Gamma}     
\newcommand{\g}{\gamma}               
\newcommand{\tg}{{\widetilde \gamma}}   
 
\newcommand{\Aintt}[1]{\int_{\Torus^d}^\oplus{#1}\,d\theta} 
\newcommand{\Zint}[1]{\int_Z^\oplus{#1}\,dz} 
\newcommand{\mR}{\mathcal R}               
\newcommand{\mB}{\mathcal B}               

\DeclareMathAlphabet{\Ma}{U}{msa}{m}{n}
\DeclareMathAlphabet{\Mb}{U}{msb}{m}{n}

\def\al #1.{{\mathcal{#1}}}

\def\mt #1.{{\mbox{\tiny $#1$}}}
\def\mrt #1.{\mathrm{\mbox{\tiny #1\,}}} 
\def\mr #1.{\mathrm{#1}}

\begin{document}

\title{Generating spectral gaps by geometry}

\author{Fernando Lled\'o}
\address{Institut f\"ur Reine und Angewandte Mathematik,
       Rheinisch-Westf\"alische Technische Hochschule Aachen,
       Templergraben 55,
       D-52062 Aachen,
       Germany}
\email{lledo@iram.rwth-aachen.de}

\author{Olaf Post}      
\email{post@iram.rwth-aachen.de}
\date{\today}



\begin{abstract}
  Motivated by the analysis of Schr\"odinger operators with periodic
  potentials we consider the following abstract situation: Let $\laplacian X$
  be the Laplacian on a non-compact Riemannian covering manifold $X$ with a
  discrete isometric group $\Gamma$ acting on it such that the quotient
  $X/\Gamma$ is a compact manifold.  We prove the existence of a finite number
  of spectral gaps for the operator $\laplacian X$ associated with a suitable
  class of manifolds $X$ with {\em non-abelian} covering transformation groups
  $\Gamma$.  This result is based on the non-abelian Floquet theory as well as
  the Min-Max-principle. Groups of type~I specify a class of examples
  satisfying the assumptions of the main theorem.
\end{abstract}
\vspace*{-0mm}
\maketitle

\section{Introduction}

It is a well known fact that a Schr\"odinger operator $-\Delta + V$ on $\R^d$,
$d\ge 2$, with a suitable periodic potential $V$ has gaps in its spectrum.
This is a quite natural situation in solid state physics, where --- for
example in insulators --- the particles described by the Schr\"odinger
operator have some unreachable energy regions (gaps). This behaviour is
ensured by the following two crucial properties: first, the fact that $V$ is
periodic.  This means that there is a basis $\{\eps_i\}_{i=1}^d$ of $\R^d$
such that the potential satisfies
\begin{displaymath}
 V(x+\eps_i) = V(x)\;,\quad i=1,\ldots,d\,.
\end{displaymath}
In other words, the periodicity of $V$ introduces an action of the discrete
abelian group $\Z^d$ on $\R^d$ and the potential is completely specified on a
fundamental domain $D \subset X$. A typical example for a fundamental domain
is the parallelepiped $D=(0,1)\eps_1 + \dots + (0,1) \eps_d$.  Second, the
potential $V$ has a high barrier near the boundary of $D$. In this way, the
potential $V$ essentially decouples the fundamental domain $D$ from the
neighbouring domains $\eps_i + D$, $i=1,\dots,d$ (see~\cite{hempel-post:03}
for an overview).

A natural question in this context is whether one can replace the effect of
the periodic potentials on the spectrum of the Laplace operator by using
geometry.  Specifically, can we replace $\R^d$ with some Riemannian manifold
$X$ with a suitable discrete group action on it, such that the corresponding
Laplace operator $\laplacian X$ also has gaps in its spectrum (which is purely
essential spectrum)? In other words, has $\spec \laplacian X$ more than one
component as a subset of $[0,\infty)$?  A positive answer to this question was
given in the context of abelian groups in \cite{post:03a} (see also the
references cited therein). The intuitive idea is that the junctions of the
fundamental domains that build up $X$ are small enough (see
Figure~\ref{fig:per-mfd} below). This has a similar effect on the energy of
the particles as the high barriers of the potential in the case of the
Schr\"odinger operator on $\R^d$. Note that the case $d=1$ is uninteresting in
our context since every Laplacian on a one-dimensional (non-compact)
Riemannian manifold is unitary equivalent to the standard one, which has no
gaps (cf.~\cite{davies-harrell:87}).  Moreover, for one-dimensional
Schr\"odinger operators any non-constant potential produces gaps
\cite[Theorem~XIII.91]{reed-simon-4}.

We will show in this paper that this simple idea of scaling down the junctions
also works for many non-abelian discrete groups $\Gamma$. The analysis in the
non-abelian case is more involved because the structure of the dual object
$\widehat \Gamma$ (i.e., the set of equivalence classes of unitary irreducible
representations of $\Gamma$) is less transparent from an algebraic and measure
theoretic point of view.  The purpose of the present paper is to stress the
fundamental ideas that allow to extend the previous result to non-abelian
group actions.  Technical details and further developments will be published
in~\cite{lledo-post:pre05}.  We will not study the nature of the spectrum
outside the gaps. Some papers related to the problem of the band-gap structure
of elliptic operators on covering manifolds are e.g.~\cite{bruening-sunada:92}
or \cite{gruber:01}. In contrast with the Schr\"odinger operator case, the
present geometric setting allows the freedom to choose the dimension $d$ of
the manifold and the number of ``period directions'' $r$ (i.e., number of
generators of $\Gamma$) \emph{independently} from each other. This observation
can probably be useful in further investigations on spectral properties common
for periodic Schr\"odinger operators and Laplacians on manifolds.

The paper is organised as follows: In the following section we set up the
problem, present the geometrical context and state some results that will be
needed later. In Section~\ref{sec:non-abelian} we introduce in detail the
non-abelian Floquet theory, which is at the basis of our analysis. We will
also illustrate the general formulas in the special case $\Gamma=\Z^d$ and
$X=\R^d$. In the next section we prove our main result: the existence of
spectral gaps in the spectrum of $\laplacian X$ for suitable manifolds $X$.
In Section~\ref{sec:examples} we specify a family of discrete groups,
so-called groups of type~I, that satisfy the assumptions of our theorem. For
convenience of the reader we have included in an appendix a short review of
the main results concerning direct integral decompositions of unitary group
representations. This technique is crucial for the Floquet theory.

\section{Notation and background}
\label{sec:prelim}
\subsection{Periodic manifolds and Laplacians}
\label{ssec:per.mfd}
We begin fixing our notation and recalling some results that will 
be useful later on. We denote by $X$ a \emph{non-compact} Riemannian
manifold of dimension $d \ge 2$. We also assume 
the action on $X$ of a finitely generated, discrete group 
$\Gamma$ of isometries of $X$ such that the
quotient $M:= X / \Gamma$ is a \emph{compact} Riemannian manifold 
which also has dimension $d$.
\begin{figure}[h]
  \begin{center}
\begin{picture}(0,0)%
\includegraphics{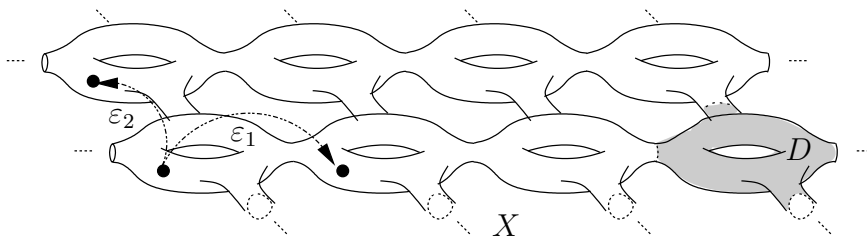}%
\end{picture}%
\setlength{\unitlength}{4144sp}%
\begin{picture}(5244,1381)(259,-575)
\put(900,120){$\eps_2$}
\put(1630,10){$\eps_1$}
\put(3196,-556){$X$}
\put(4950,-100){$D$}
\end{picture}
    \caption{A periodic (or covering) manifold $X$ with group $\Gamma$
      generated by two elements $\eps_1, \eps_2$ and fundamental domain $D$
      (in grey). Here, the group $\Gamma$ is abelian.}
    \label{fig:per-mfd}
  \end{center}
\end{figure}
In other words, $X$ is a \emph{periodic manifold} or \emph{Riemannian covering
  space} of $M$ with covering transformation group $\Gamma$. Moreover, we fix
a \emph{fundamental domain} $D$, i.e., an open set $D \subset X$ such that $\g
D$ and $\g' D$ are disjoint for all $\g \ne \g'$ and $\bigcup_{\g \in \G} \g
\overline D = X$ (cf.\ Figure~\ref{fig:per-mfd}).

As a prototype for an elliptic operator we consider the Laplacian $\laplacian
X$ on $X$ acting on a dense subspace of the Hilbert space $\Lsqr X$
with norm $\norm[X] \cdot$. The positive self-adjoint operator 
$\laplacian X$ can be defined in terms of a suitable 
quadratic form $q_X$ (see e.g.~\cite[Chapter~VI]{kato:95},
\cite{reed-simon-1} or \cite{davies:96}). Concretely we have
\begin{equation}
\label{def:quad.form}
  q_X(u):=\normsqr[X] {d u} = \int_{X} |d u|^2 dX\;,\quad
          u \in \Cci X \;.
\end{equation}
In coordinates we write the pointwise norm of the $1$-form $d u$ as
\begin{displaymath}
  |d u |^2= \sum_{i,j} g^{ij} \partial_i u \, \partial_j\overline u \,,
\end{displaymath}
where $(g^{ij})$ is the inverse of the metric tensor $(g_{ij})$ in a
chart. Taking the closure of the quadratic form we can extend $q_X$ onto
the Sobolev space
\begin{displaymath}
  \Sob X = \set{ u \in \Lsqr X}{ q_X(u) < \infty}\,.
\end{displaymath}
As usual the operator $\laplacian X$ is related with the quadratic form by the
formula $\iprod {\laplacian X u} u = q_X(u)$, $u \in \Cci X$. Since the metric
on $X$ is $\Gamma$-invariant, the Laplacian $\laplacian X$ (i.e., its
resolvent) commutes with the translations on $X$ given by
\begin{equation}
  \label{eq:transl}
   (T_\gamma u )(x) := u(\gamma^{-1}x), \quad u \in \Lsqr X, \gamma \in \Gamma.
\end{equation}
Operators with this property are called \emph{periodic}.

For an open, relatively compact subset $D \subset X$ with sufficiently smooth
boundary $\bd D$ (e.g.~Lipschitz) we define the Dirichlet (respectively,
Neumann) Laplacian $\laplacianD D$ (resp., $\laplacianN D$) via its quadratic
form $q_D^+$ (resp., $q_D^-$) associated to the closure of $q_D$ on $\Cci D$,
the space of smooth functions with compact support, (resp.\, $\Ci {\overline
  D}$, the space of smooth functions with continuous derivatives up to the
boundary).  We also use the notation $\Sobn D = \dom q_D^+$ (resp., $\Sob D =
\dom q_D^-$). Note that the usual boundary condition of the Neumann Laplacian
occurs only in the \emph{operator} domain via the Gau{\ss}-Green formula.
Since $\overline D$ is compact, $\laplacianD D$ has purely discrete spectrum
$\EWD k$, $k \in \N$.  It is written in ascending order and repeated according
to multiplicity.  The same is true for the Neumann Laplacian and we denote the
corresponding purely discrete spectrum by $\EWN k$, $k\in\N$.

One of the advantages of the quadratic form approach is that 
one can easily read off from the inclusion of domains an order relation 
for the eigenvalues. In fact, by the 
the \emph{min-max principle} we have
\begin{equation}
  \label{eq:min.max}
  \EWDN k =
  \inf_{L_k} \sup_{u \in L_k \setminus \{0\} }
      \frac {q_D^\pm(u)}{\normsqr u}\;,
\end{equation}
where the infimum is taken over all $k$-dimensional subspaces $L_k$ of the
corresponding \emph{quadratic} form domain $\dom q_D^\pm$. Then the 
inclusion
\begin{equation}
  \label{eq:dom.mono}
   \dom q_D^+ = \Sobn D  \subset \Sob D = \dom q_D^- 
\end{equation}
implies $ \EWD k \ge \EWN k$, i.e., the Dirichlet eigenvalue is in general
larger than the Neumann eigenvalue and this justifies the choice of the 
labels $+$, respectively, $-$.

\subsection{Spectral gaps in the abelian case}
\label{sec:gaps.abel}
Due to the previous inequality relating the Dirichlet and Neumann 
eigenvalues for any $k\in\N$, we may introduce the following intervals
\begin{equation}
  \label{eq:def.Ik}
  I_k := [\EWN k, \EWD k]\;,\quad k\in\N\,.
\end{equation}
Here, $\EWDN k$ denotes the $k$-th Dirichlet/Neumann eigenvalue on a
fundamental domain $D$.

In this context the existence of spectral gaps of $\laplacian X$ can be
reduced to the question whether $I_k \cap I_{k+1} = \emptyset$ for some $k$.
A class of manifolds with abelian group actions satisfying the previous
intersection condition is specified in \cite{post:03a}:
\begin{theorem}
  \label{thm:ex.gaps.ab}
  Given a finitely generated abelian group $\Gamma$ we can always find a
  covering space $X \to X/\Gamma =: M$ and a fundamental domain $D$ with the
  following property: For every $K \in \N$ there exists a metric $g=g_K$ on
  $M$ such that
  \begin{equation}
    \label{eq:gaps}
    I_k \cap I_{k+1} = \emptyset
  \end{equation}
  for at least $K$ indices $k \in \N$, where $I_k$ is defined as in
  \eqref{eq:def.Ik} for the manifold $D$ with metric $g=g_K$. In particular,
  the Laplacian $\laplacian X$ corresponding to the lifted metric on $X$ has
  at least $K$ gaps.
\end{theorem}
It is important to note that the construction of the covering space $X$ and
the metric $g$ only depends on the quotient $M$, not on the cover $X$. It is
therefore independent whether $\Gamma$ is abelian or not (see the sketch of
the proof of Theorem~\ref{thm:ex.gaps}).  Roughly speaking, we have replaced
the \emph{high} potential barrier in the case of Schr\"odinger operators on
$\R^d$ by \emph{small} junctions between the fundamental domains.  We can say
that now geometry is partly decoupling one fundamental domain from its
neighbours.

\section{Non-abelian Floquet theory}
\label{sec:non-abelian}

In this section we will introduce in several steps the Floquet theory for
non-abelian groups. The main idea is to use the group action on $X$ and a
partial Fourier transformation to decompose the Hilbert space $\Lsqr X$ and
the periodic operators on it into a direct integral of simpler components that
can be analysed more easily. For convenience of the reader we have summarized
in the appendix the main results on the direct integral decompositions of
unitary group representations.

\subsection{Non-abelian Fourier transformation}
\label{ssec:non-ab.fourier}

Consider first the right, respectively, left regular representation $R$,
resp., $L$ on the Hilbert space $\lsqr \G$:
\begin{equation}
  \label{def:reg.rep}
  (R_\g a)_\tg = a_{\tg \g}, \qquad
  (L_\g a)_\tg = a_{\g^{-1}\tg},  \quad\qquad a = (a_\g)_\g \in \lsqr \G, 
      \quad \g,\tg \in \G.
\end{equation}
Let $\mR$ be the von Neumann algebra generated by all unitaries $R_\g$, $\g
\in \G$, i.e.,
\begin{displaymath}
  \mR = \set {R_\g} {\g \in \G}'',
\end{displaymath}
and denote by $\mR'$ the commutant of $\mR$ in $\mB(\lsqr \G)$;
similarly, we define $\mathcal L = \set {L_\g} {\g \in \G}''$.

In this context, we may generalise the Fourier transformation to the
unitary map
\begin{equation}
  \label{eq:fourier}
  \map F {\lsqr \G}{\Zint {H(z)}}
\end{equation}
that transforms the right regular representation $R$ into the following direct
integral representation
\begin{equation}
  \label{eq:reg.rep.trafo}
    \widehat R_\g = F R_\g F^{-1} = \Zint {R_\g(z)}, \qquad \g \in \G.
\end{equation}
By a suitable choice of the measure space $(Z,dz)$ corresponding to an
maximal abelian algebra $\mathcal A$ in $\mathcal R'$ (see the
appendix) we can assume that the unitary representations $R_\g(z)$ are
irreducible on the Hilbert space $H(z)$ a.e. In addition, operators
commuting with all $L_\gamma$ ($\gamma \in \Gamma$), i.e., operators in
$\mathcal L'$, are decomposable, since one can show that $\mathcal
L'=\mathcal R$ and therefore $\mathcal L' \subset \mathcal A'$ (cf.\ 
the appendix).

\subsection{Equivariant Laplacians}
\label{ssec:equiv.lapl}
We will introduce next a new operator that lies ``between''
the Dirichlet and Neumann Laplacians and that will play an important
role in the following section. Consider
on almost each fibre smooth \emph{$R(z)$-equivariant
  functions}, i.e., smooth functions $\map h X {H(z)}$ satisfying
\begin{equation}
  \label{def:equiv.fct}
    h(\g x) = R_\g(z) h(x), \qquad  \g \in \G, x \in X.
\end{equation}  
We denote the corresponding space of smooth $R(z)$-equivariant functions
\emph{restricted to a fundamental domain $D$} by $\CiZ {D,H(z)}$. Note that we
need \emph{vector-valued} functions $\map h X {H(z)}$ since the representation
$R(z)$ acts on the Hilbert space $H(z)$. If $\Gamma$ is non-abelian, then some
of its unitary irreducible representations must be of dimension greater than
one.

We introduce next the so-called \emph{equivariant Laplacian} (w.r.t.\ 
the representation $R(z)$) on $\Lsqr {D,H(z)} \cong \Lsqr D \otimes
H(z)$. Consider the quadratic form given by
\begin{equation}
\label{eq:quad.form}
  \normsqr[D] {dh} := \int_D \normsqr[H(z)] {dh(x)} dX
\end{equation}
for $h \in \CiZ {D,H(z)}$, where the integrand is locally specified by
\begin{displaymath}
    \normsqr[H(z)] {dh(x)} = 
    \sum_{i,j} g^{ij}(x) \, \iprod[H(z)] {\partial_i h(x)} {\partial_j h(x)},
     \qquad x \in D.
\end{displaymath}
This generalises Eq.~\eqref{def:quad.form} to the case of vector-valued
functions. We denote the closure by $q_D^{\mrt eq.}$ and its domain by
$\SobZ{D,H(z)}$. The corresponding non-negative operator on $\Lsqr {D,H(z)}$,
the so-called \emph{$R(z)$-equivariant Laplacian}, will be denoted by
$\laplacianZ D$.

\subsection{Non-abelian Floquet transformation}
\label{ssec:non-ab.floquet}
Next, we analyse the Floquet transformation
\begin{displaymath}
  \map U {\Lsqr X} {\Zint {\Lsqr {D,H(z)}}},
\end{displaymath}
which is the composition of the following three unitary transformations
(denoted with horizontal arrows)
\begin{displaymath}
  \begin{array}{ccr@{\otimes}lcr@{\otimes}lcc}
    \stackrel{T_\g} \curvearrowright & & 
    \multicolumn{2}{c}{\stackrel{L_\g \otimes \1} \curvearrowright \quad} && 
    \multicolumn{2}{c}{\stackrel{\widehat L_\g \otimes \1} 
                                                    \curvearrowright \quad} &&
    \\ 
    \Lsqr X & \!\!\!\!\to\!\!\!\! &        
    \lsqr \G & \Lsqr D & \!\!\stackrel {F \otimes \id} \to\!\! & 
    \Zint {H(z)} & \Lsqr D & \!\!\!\!\to\!\!\!\!&
    \Zint {\Lsqr {D,H(z)}}\\[5ex]
    u & \mapsto &                      
    \sum_\g \delta_\g & (T_{\g^{-1}} u \restr D) & &
    b & f & \mapsto & 
    (b(\rho)f)_\rho \;,
  \end{array} 
\end{displaymath}
where $(\delta_\gamma)_\gamma$ is the canonical orthogonal basis of
$\lsqr \G$ and $T_\gamma$ is the translation by $\gamma$ of functions
on $X$ given by~\eqref{eq:transl}.  Each of these transformations
intertwines with the unitary representation of $\G$ which are denoted
with curved arrows in the previous diagram.  The first horizontal
unitary transformation just splits a function on $\Lsqr X$ into a
sequence of $\g$-translates over the fundamental domain $D$. The
second horizontal unitary is essentially the Fourier transformation on
the group part and the last horizontal unitary is clear, since $\Lsqr
D$ is independent of $z\in Z$.

Note that periodic operators on $\Lsqr X$, i.e, operators commuting
with $T_\gamma$, are those commuting with $L_\gamma \otimes \1$ on
$\lsqr \G \otimes \Lsqr D$. Therefore, periodic operators are also
decomposable (recall that $\mathcal L' = \mathcal R \subset \mathcal
A'$).  Note in addition that if $\Gamma$ is not abelian then $\widehat
L_\gamma = F L_\gamma F^{-1}$ does not decompose with respect to the
direct integral specified in Section~\ref{ssec:non-ab.fourier}.

The Floquet transformation $U$ is given explicitely in the following theorem
(see also~\cite{sunada:88} and~\cite[Section~XIII.16]{reed-simon-4}):
\begin{theorem}
  \label{thm:floquet}
  The map
  \begin{equation}
    \label{eq:floquet.short}
    (Uu)(z)(x) = \sum_{\g \in \G} \,u(\gamma x)\;  R_{\g^{-1}}(z) v(z),
    \quad \text{where $v := F \delta_e$,}
  \end{equation}
  is a unitary transformation that intertwines the representations $T$ and
  $\Zint {R(z)}$. In addition, $U$ maps $\Cci X$ into $\Zint {\CiZ {D,H(z)}}$
  and operators on $\Lsqr X$ commuting with all $T_\g$'s are decomposable when
  transformed onto the direct integral.  In particular, the Laplacian
  $\laplacian X$ is unitary equivalent to $\Zint {\laplacianZ D}$ and
  \begin{equation}
  \label{eq:spec.dir.int}
    \spec \laplacian X \subseteq
    \overline {\bigcup_{z \in Z} \spec \laplacianZ D}.
  \end{equation}
\end{theorem}


\subsection{The special case $\Gamma=\Z^d$ and $X=\R^d$.}
\label{ssec:z.d}
In this case the dual is simply given by the $d$-dimensional torus, i.e., $\hG
= \Torus^d$. Therefore, we can choose $Z = \Torus^d$ with Lebesgue measure
$d\theta$.  The Fourier transformation~\eqref{eq:fourier} reduces to the
standard formula
\begin{displaymath}
    \map F {\lsqr {\Z^d}} {\Lsqr {\Torus^d} = \Aintt {H(\theta)}}\;, \quad
    \mr with.~(Fa)(\theta) = \sum_\g \, \e^{-\im  \theta \cdot \g} a_\g
\end{displaymath}
where $H(\theta)=\C$ for a.e.~$\theta \in \Torus^d = \R^d/\Z^d$ and
$a=(a_\g) \in \lsqr {\Z^d}$.  Here, we only need scalar functions
since every irreducible representation of an \emph{abelian} group is
one-dimensional.  In this case, we can decompose \emph{both} the left
\emph{and} the right regular representation simultaneously; each fibre
of $R$, resp., $L$ is the multiplication with a phase factor
\begin{displaymath}
  \widehat R_\g(\theta) = 
               \e^{\im \theta \cdot \g}, 
     \quad \text{resp.,} \quad
  \widehat L_\g(\theta)  = 
               \e^{-\im \theta \cdot \g} 
\end{displaymath}
 The equivariant condition becomes
\begin{equation}
\label{def:equiv.fct.ab}
   h(x+\gamma) = \e^{-\im \theta \cdot \gamma} h(x)
\end{equation}
for all $x \in \R^d$ and $\gamma \in \Z^d$. A fundamental domain is the cube
$D := (0,1)^d$ and the Floquet transformation is given by
\begin{displaymath}
  (Uu)(\theta)(x) = 
     \sum_{\gamma \in \Z^d} u(x+\gamma) \e^{\im \theta \cdot \gamma}, 
        \qquad x \in D, \theta \in \Torus^d.
\end{displaymath}

\section{Existence of spectral gaps for non-abelian group actions}
\label{sec:main.result}
We will present in this section a method to show the existence of finitely
many spectral gaps for the Laplace operator $\laplacian X$ in the case of
\emph{non-abelian} group actions. Our main assumption on the group is the fact
that the irreducible representation appearing in the
decomposition~\eqref{eq:fourier} are finite-dimensional a.e., in other words
\begin{equation}
  \label{eq:dim.finite}
  \dim H(z) < \infty \qquad \text{for a.e.\ $z \in Z$}.
\end{equation}

The operators $\laplacianD D(z)$, $\laplacianZ D$, resp., $\laplacianN D(z)$
corresponding to the quadratic form~\eqref{eq:quad.form} on $\Sobn {D,H(z)}$,
$\SobZ {D,H(z)}$, resp., $\Sob {D,H(z)}$ have purely discrete spectrum which
we denote by $\EWD m(z)$, $\EWZ m$, resp., $\EWN m (z)$, $m \in \N$.  Recall
that the space $\Sobn {D,H(z)}$ is the $\HS^1$-closure of the space of smooth
functions $\map h D {H(z)}$ with compact support and $\Sob {D,H(z)}$ is the
closure of the space of smooth functions with derivatives continuous up to the
boundary (cf.~Section~\ref{sec:prelim}).

As in \eqref{eq:dom.mono} we obtain from the inclusion of the three
domains
\begin{displaymath}
  \Sob {D,H(z)} \supset \SobZ {D,H(z)} \supset \Sobn {D,H(z)}
\end{displaymath}
that the corresponding eigenvalues satisfy the following reverse inequalities
\begin{displaymath}
 \EWN m (z)  \le \EWZ m  \le \EWD m (z)
\end{displaymath}
for all $m \in \N$ and a.e.\ $z \in Z$.

From the definition of the quadratic form in the Dirichlet, resp., Neumann case
we have that the corresponding vector-valued Laplacians are a direct sum of
the scalar operators since there is no coupling between the components on the
boundary. In particular, if $n = \dim H(z)$, then $\laplacianDN D (z)$ is a
$n$-fold direct sum of the scalar operators $\laplacianDN D$ on $\Lsqr D$.
Therefore the eigenvalues of the corresponding vector-valued Laplace operators
consist of $n$-times repeated eigenvalues of the scalar Laplacians. We can
therefore arrange the former in the following way:
\begin{displaymath}
  \EWDN m (z) = \EWDN k, \qquad
  m=(k-1)n+1, \dots, kn\,,
\end{displaymath}
where $\EWDN k$ denotes the (scalar) $k$-th Dirichlet/Neumann eigenvalue on
$D$.

Recall the definition of the intervals $I_k := [\EWN k, \EWD k]$ in
Eq.~(\ref{eq:def.Ik}).  We may now collect the $n$ eigenvalues
of $\laplacianZ D$ which lie in $I_k$:
\begin{equation}
  \label{eq:band.z}
  B_k(z) := \set {\EWZ m} 
                  {m=(k-1)n+1, \dots, kn}
          \subset I_k,  \qquad n := \dim H(z)\,.
\end{equation}
Moreover we put together all eigenvalues 
corresponding to operators over the base point $z \in Z$ 
that act on Hilbert spaces with the same dimension:
\begin{equation}
  \label{eq:band.r}
    B_k(n) := \bigcup_{z \in Z, \, \dim H(z) = n} B_k(z) \subset I_k.
\end{equation}

\begin{theorem}
  \label{thm:spec.incl}
  Let $\Gamma$ be a finitely generated (in general non-abelian) group  
  satisfying $\dim H(z) < \infty$ for a.e.\ $z \in Z$ in the 
  decomposition~\eqref{eq:fourier}. Then
  \begin{displaymath}
    \spec \laplacian X \subseteq \bigcup_{k \in \N} I_k.
  \end{displaymath}
  In particular, the spectrum of $\laplacian X$ has a gap between $I_k$ and
  $I_{k+1}$ provided 
  \begin{displaymath}
         I_k \cap I_{k+1} = \emptyset\;.
  \end{displaymath}
\end{theorem}
\begin{proof}
  The proof is a consequence of the following chain of inclusions
  \begin{equation}
    \label{eq:spec.incl2}
    \spec \laplacian X \subseteq
    \overline{\bigcup_{z \in Z} \spec \laplacianZ D} =
    \overline{\bigcup_{z \in Z} \bigcup_{k \in \N} B_k(z)} =
    \overline{\bigcup_{n \in \N} \bigcup_{k \in \N} B_k(n)} \subseteq
    \bigcup_{k \in \N} I_k\,.
  \end{equation}
  For the first inclusion, we have applied~\eqref{eq:spec.dir.int}. For the
  second equality we have used $\dim H(z) < \infty$. Note finally that
  $\bigcup_k I_k$ is closed since $\EWDN k \to \infty$ for $k \to \infty$
  implies that $(I_k)_k$ is a locally finite family of closed sets.
\end{proof}

We formulate our main result, i.e., an analogue of
Theorem~\ref{thm:ex.gaps.ab} in the case of non-abelian groups $\Gamma$
satisfying the conditions stated in Section~\ref{sec:prelim}:

\begin{theorem}
  \label{thm:ex.gaps}
  Let $\Gamma$ be a finitely generated (in general non-abelian) group
  satisfying $\dim H(z) < \infty$ for a.e.\ $z \in Z$ in the
  decomposition~\eqref{eq:fourier}. Then we can always find a covering space
  $X \to X/\Gamma$ and a fundamental domain $D$ with the following property:
  For each $K\in\N$ there exist a Riemannian covering space $X \to
  X/\Gamma=:M$ with metric $g=g_K$ such that the Laplacian on $X$ has at least
  $K$ gaps, i.e., $\spec \laplacian X \subset [0, \infty)$ has at least $K$
  components.
\end{theorem}
\begin{proof}[Sketch of the proof:]
  Suppose that $\Gamma$ has $r$ generators $\varepsilon_i$.  To construct $X$
  we begin with a $d$-dimensional compact manifold $M$ with at least $r$
  handles which we may take diffeomorphic to $(0,1)\times\Sphere^{d-1}$. Then
  $D\subset M$ is the open subset obtained from $M$ by removing a section
  $\{1/2\} \times \Sphere^{d-1}$ from each of the $r$ handles. The set $D$ has
  therefore $2r$ cylindrical ends, for each generator a ``left'' and a
  ``right'' one.  Then there exists a covering $X \to X/\Gamma \cong M$ with
  fundamental domain $D$:  Intuitively, one can build up $X$ by glueing
  $\Gamma$ copies $(\gamma D)_\gamma$ of $D$, where one has to identify
  properly the points on the boundary (cf.~Fig.~\ref{fig:per-mfd} in the case
  $r=2$).  Concretely, we identify the ``left'' boundary part of the $i$-th
  cylindrical end of $\gamma_1 D$ with the ``right'' one of $\gamma_2 D$ iff
  $\gamma_2 = \eps_i \gamma_1$.  Finally, we change the metric on the handles
  in order to scale down the junctions between neighbouring copies of the
  fundamental domain (cf.~\cite{post:03a}). Note that the metric depends on
  the minimal number $K$ of gaps.  This implies that Eq.~\eqref{eq:gaps} is
  satisfied for at least $K$ indices $k \in \N$ and the proof is concluded by
  Theorem~\ref{thm:spec.incl}.
  
  Note that we have spectrum in each interval $I_k$, i.e., $\spec
  \laplacian X \cap I_k \ne \emptyset$ for all $k$, since a group
  satisfying~\eqref{eq:dim.finite} is amenable (cf.~the next section);
  therefore, $\spec \laplacian M \subset \spec \laplacian X$ (see
  e.g.~\cite[Prop.~7--8]{sunada:88}). Finally, $\spec \laplacian M
  \cap I_k \ne \emptyset$.
\end{proof}

Note that the previous statement does not give information on the
\emph{maximal} number of spectral gaps.  It still remains an open question if
there are (connected) covering spaces $X$ with an \emph{infinite} number of
spectral gaps.  This problem is related to the so-called
\emph{Bethe-Sommerfeld conjecture} (cf.~\cite{skriganov:87}).

\section{Examples}
\label{sec:examples}

We begin defining a class of a discrete groups that have particularly simple
properties (cf.~\cite{thoma:64}). A discrete group $\Gamma$ is of type~I iff
there is an exact sequence
\begin{displaymath}
  0 \longrightarrow 
  A \longrightarrow
  \G \longrightarrow
  \G_0 \longrightarrow 
  0,
\end{displaymath}
where $A$ is a finitely generated \emph{abelian} normal subgroup of $\Gamma$
and $\G_0 = \G / A$ is a \emph{finite} group (cf.~\cite{thoma:64}). Simple
examples of groups of type~I are abelian groups (in this case $\Gamma_0$ is
trivial) or direct, resp., semi-direct products of an abelian group with a
finite (in general non-abelian) group.

For these type of groups we have that all irreducible representations
are finite dimensional. In addition, such groups are also amenable (as
extensions of amenable groups), cf.~\cite{brooks:81}. Note that the
converse is also true: A discrete group such that all irreducible
representations are finite dimensional is of type~I
(cf.~\cite{moore:72}).

Recall that this was an important assumption in
Section~\ref{sec:main.result}.  Moreover in the direct integral decomposition
of Subsection~\ref{ssec:non-ab.floquet} we may take as measure space $Z$ the
dual $\hG$ of $\Gamma$. For these reasons we have

\begin{proposition}
  \label{thm:groups.type.I}
  Suppose $\G$ is a finitely generated group of type~I.  Then for each $K \in
  \N$ there exist a Riemannian covering space $X \to X/\Gamma$ with metric
  $g=g_K$ such that the Laplacian on $X$ has at least $K$ gaps, i.e., $\spec
  \laplacian X \subset [0, \infty)$ has at least $K$ components.
\end{proposition}
Of course, not all groups are of type~I. For example, free groups with more
than one generator are not of type~I. In~\cite{lledo-post:pre05} we provide
different methods and further classes of groups including the free groups for
which the conclusions of the above proposition remain true.

\section*{Appendix: Direct integral decomposition of unitary 
          group representations}

In the present appendix we will describe in more detail the direct integral
decomposition of the right regular representation given in
Eqs.~\eqref{def:reg.rep} and \eqref{eq:fourier} of
Section~\ref{sec:non-abelian}. It is an application of the direct integral
decomposition of von Neumann algebras. General references are
e.g.~\cite[Chapter~14]{wallach:92} or~\cite[Chapter~2]{mackey:76}.

In this appendix we will consider a more general frame that includes the
particular situation considered in Eqs.~\eqref{eq:fourier} and
\eqref{eq:reg.rep.trafo}, where $\Gamma$ is a discrete group satisfying the
conditions of Section~\ref{sec:prelim} and $R$ is the right regular
representation on $\ell_2(\Gamma)$.

Let $G$ be a separable locally compact group and let $V$ be a continuous
unitary representation of $G$ on a Hilbert space $\HS$. Denote by
\begin{displaymath}
  \al M.:=\{V_g \mid g\in G\}'' 
\end{displaymath}
the von Neumann algebra generated by the representation $V$
and let
\begin{displaymath}
 \al M.'=(V,V):=\{M'\in\al B.(\al H.)\mid M'  V_g=V_g  M'\;,\;g\in G\} 
\end{displaymath}
be the von Neumann algebra of operators commuting with the representation $V$.
If $\al A.$ is an abelian von Neumann subalgebra of $\al M.'$, then there
exists a compact, separable Hausdorff space $Z$, a regular Borel measure $dz$
on $Z$ and a unitary transformation onto a direct integral Hilbert space
\begin{equation}
  \label{eq:dir.int}\tag{A.1}
  \map F {\al H.}{\Zint {\al H.(z)}} ,
\end{equation}
such that
\begin{displaymath}
   F \al A. F^{-1}=\set {M_f} {f\in L_\infty(Z,dz)}
\end{displaymath}
($M_f$ being the multiplication operator with $f$). The von Neumann
algebra $F \mathcal A' F^{-1}$ consists of all decomposable operators
w.r.t.\ the direct integral~\eqref{eq:dir.int}, i.e., if $D \in F
\mathcal A' F^{-1}$ then we can write
\begin{displaymath}
  D = \Zint {D(z)}.
\end{displaymath}
In particular, $V_g \in \mathcal M \subset \mathcal A'$, and therefore
\begin{displaymath}
  F  V_g  F^{-1}= \Zint {V_g(z)},
\end{displaymath}
where $V(z)$ is a unitary representation of $G$ on $H(z)$ a.e. (see
\cite[Section~14.8~ff.]{wallach:92}). There are several natural choices for
the abelian von Neumann algebra $\al A.$:
\begin{itemize}
\item[(i)] If $\al A.=\al M.\cap\al M.'$ is the centre of $\al M.$, then, for
  a.e.~$z\in Z$, the von Neumann algebra generated by the representations
  $V(z)$ are factors, i.e.,
\begin{displaymath}
 \al M.(z)\cap\al M.(z)'
    :=\{V_g(z)\mid g\in G\}''\cap\{V_g(z)\mid g\in G\}'=\C\1_{\al H.(z)}.
\end{displaymath} 
This choice is due to von Neumann.

\item[(ii)] If $\al A.$ is maximal abelian in $\al M.'$, i.e., $\al A.=\al
  A.'\cap\al M.'$, then the components $V(z)$ of the direct integral
  decomposition of $V$ are irreducible a.e.  This choice is due to Mautner and
  was used in Eqs.~\eqref{eq:fourier} and~\eqref{eq:reg.rep.trafo} of
  Section~\ref{sec:non-abelian}.
\end{itemize}

Finally, we mention a class of groups, where the previous decomposition
results become particularly simple.  A group $G$ is of type~I if all its
unitary continuous representations $V$ are of type~I, i.e., each $V$ is
quasi-equivalent to some multiplicity free representation. Compact or abelian
groups are examples of type~I groups. If $G$ is of type~I, then the dual
$\widehat{G}$ (i.e., the set of all equivalence classes of continuous unitary
irreducible representations of $G$) becomes a nice measure space (``smooth''
in the terminology of \cite[Chapter~2]{mackey:76}).  In this case one can take
$\widehat{G}$ as the measure space $Z$ in the Mautner decomposition (ii).  For
discrete groups the previous definition of Type~I is equivalent to the one
given in Section~\ref{sec:examples} (cf.~\cite{thoma:64}).

\section*{Acknowledgements}
The second author would like to thank the organisers of the Young Researchers
Symposium at the Instituto Superior T\'ecnico (Lisbon) in July 2003 for the
kind invitation. We would also like to thank Volker En{\ss} for useful remarks
on the manuscript.




\end{document}